\documentclass[aps,pra,twocolumn,superscriptaddress,groupedaddress]{revtex4}

\usepackage{amssymb} \usepackage{color,graphicx} \usepackage{amsmath}
\usepackage{amsbsy} \usepackage{amsthm} \usepackage{bbm}
\usepackage{bm,bbm} \usepackage{float} \usepackage{braket}
\usepackage{placeins}
\usepackage[colorlinks=true,citecolor=blue,linkcolor=red,urlcolor=red]{hyperref}
\usepackage[sort&compress]{natbib}
\usepackage{comment,cprotect}
\usepackage{stackengine}

\usepackage{array}
\newcolumntype{C}{>{$\displaystyle} c <{$}}

\newcommand{\bq}{\begin{equation}} \newcommand{\eq}{\end{equation}}
\newcommand{\bqali}{\bq\begin{aligned}}
\newcommand{\eqali}{\end{aligned}\eq}
\newcommand{\bqn}{\begin{equation*}}
\newcommand{\eqn}{\end{equation*}}

\newcommand\D{\operatorname{d}\!}
\renewcommand\k{{\bf k}}

\newcommand\p{{\bf p}}
\newcommand\n{{\bf n}}
\newcommand\z{{\bf z}}

\newcommand\q{{\bf q}}

\newcommand\x{{\bf x}}
\newcommand\y{{\bf y}}

\newcommand\com[2]{[#1,#2]}

\newcommand\rC{r_\text{\tiny C}}

\begin{document}

\author{Stephen L. Adler}
\affiliation{Institute for Advanced Study, Einstein Drive, Princeton, New Jersey 08540, USA}
\author{Angelo Bassi}
\affiliation{Department of Physics, University of Trieste, Strada Costiera 11, 34151 Trieste, Italy}
\affiliation{Istituto Nazionale di Fisica Nucleare, Trieste Section, Via Valerio 2, 34127 Trieste, Italy}
\author{Matteo Carlesso}
\email{matteo.carlesso@ts.infn.it}
\affiliation{Department of Physics, University of Trieste, Strada Costiera 11, 34151 Trieste, Italy}
\affiliation{Istituto Nazionale di Fisica Nucleare, Trieste Section, Via Valerio 2, 34127 Trieste, Italy}
\author{Andrea Vinante}
\affiliation{Department of Physics and Astronomy, University of Southampton, SO17 1BJ, United Kingdom}

\title{Testing Continuous Spontaneous Localization with Fermi liquids}

\date{\today}
\begin{abstract}

{
Collapse models describe phenomenologically the quantum-to-classical transition by adding suitable nonlinear and stochastic terms to the Schr\"odinger equation, thus (slightly) modifying the dynamics of quantum systems. Experimental bounds on the collapse parameters have been derived from various experiments involving a plethora of different systems, from single atoms to gravitational wave detectors. Here, we give a comprehensive treatment of the Continuous Spontaneous Localization (CSL) model, the most studied among collapse models, for Fermi liquids. We consider both the white and non-white noise case. Application to various astrophysical sources is presented.}

\end{abstract}
%
 \maketitle

\section{Introduction}

Collapse models provide a phenomenological description of quantum measurements,  by adding stochastic and non-linear terms to the Schr\"odinger equation, which implement the collapse of the  wave function~\cite{Bassi:2003aa}. Such effects are negligible for microscopic systems, and become stronger when their mass increases. This is how the quantum-to-classical transition  is described {}{and the measurement problem solved, which is the main motivation why they were formulated in the first place}.

The most studied model is the Continuous Spontaneous Localization (CSL) model \cite{Pearle:1989aa, Ghirardi:1990aa}. {}{It applies to identical particles and the collapse, which is implemented by a noise  coupled nonlinearly to the mass-density of the system, occurs continuously in time.}  The collapse effects are quantified by two parameters: the collapse rate $\lambda$, and the correlation length of the  noise $\rC$. Different theoretical proposals for their numerical value were suggested: $\lambda=10^{-16}\,$s$^{-1}$ and $\rC=10^{-7}\,$m by Ghirardi, Rimini and Weber \cite{Ghirardi:1986aa}; $\lambda=10^{-8\pm2}\,$s$^{-1}$ for $\rC=10^{-7}\,$m, and $\lambda=10^{-6\pm2}\,$s$^{-1}$ for $\rC=10^{-6}\,$m by Adler \cite{Adler:2007ab}.
Experimental data were extensively used to bound the parameters \cite{Adler:2007ab,Adler:2007ad,Adler:2013aa,Donadi:2013aa,Bahrami:2013aa,Donadi:2013ab,Bassi:2014aa,Donadi:2014aa,Belli:2016aa,Bilardello:2016aa,Curceanu:2016aa,Vinante:2016aa,Carlesso:2016ac,Helou:2017aa,Vinante:2017aa,Toros:2017aa,Piscicchia:2017aa,Toros:2018aa,Adler:2018aa,Carlesso:2018ab} and new proposals were presented, suggesting how to further push these bounds \cite{Collett:2003aa,Goldwater:2016aa,Kaltenbaek:2016aa,McMillen:2017aa,Carlesso:2018ac,Schrinski:2017aa,Carlesso:2018ab,Mishra:2018aa}. Fig.~\ref{fig1} summarizes the state of the art.

{}{In this context, one important question is  the origin of the collapse noise. While collapse models do not give an answer, as the  collapse is  inserted `by hand' into the Schr\"odinger dynamics (but its mathematical structure is fully constrained by the request of no-superluminal-singling and norm-conservation \cite{Bassi:2003aa}), several times it has been suggested that is related to gravity \cite{Diosi:1984aa,Diosi:1987aa, Diosi:1989aa,Penrose:1996aa,Pearle:1996aa,Diosi:2007aa,Giulini:2012aa,Tilloy:2016aa,Adler:2016ab,Gasbarri:2017aa,Tilloy:2018aa}. If there is truth in this conjecture, then the gravitational fluctuations responsible for the collapse add to the usual gravitational effects present in matter, in particular in strongly gravitationally bound systems as those we will consider in this paper.}

{A consequence of collapse models is a spontaneous heating, induced by the random collapse.  This effect 
has been calculated for many types of systems \cite{Belli:2016aa,Bilardello:2016aa,Vinante:2016aa,Carlesso:2016ac,Helou:2017aa,Vinante:2017aa,Adler:2018aa,Carlesso:2018ab}, but not for Fermi liquids, an issue raised in a recent 
paper of Tilloy and Stace \cite{Tilloy:2019aa}.   Here, we give a comprehensive treatment of CSL induced heating in Fermi 
liquids, including the experimentally relevant case of non-white noise,  and apply our results to various 
astrophysical systems, including neutron stars.  }

\section{CSL model - perturbative calculation}

Following \cite{Adler:2018aa}, we consider the transition amplitude $c_{fi}(t)$ caused by a perturbation, from an initial state $\ket i$ of a quantum system to a final state $\ket f$, with associated energies $E_i=\hbar \omega_i$ and $E_f=\hbar \omega_f$ respectively. For the sake of simplicity we restrict the problem to the case of one {fermion} of mass $m_A$. The result for the $N$ particle case{, either fermions or bosons,} is given  in Appendix \ref{app.N}. We have:
\bq
c_{fi}(t)=-\frac i\hbar\int_0^t\D s\,\braket{f|e^{\tfrac i\hbar\hat H_0s}\hat V(s)e^{-\tfrac i\hbar\hat H_0s}|i},
\eq
where $\hat H_0$ is the free Hamiltonian  and the perturbation, for the CSL process applied to a particle of mass $m_A$, is \cite{Adler:2018aa}:
\bqali
\hat V(t)&=\int\D\z\,w_t(\z)\hat{\mathcal V}(\z),\\
\hat{\mathcal V}(\z)&=-\frac{\hbar}{m_0}m_A g(\z-\hat \x_A),
\eqali
where $m_0$ is the nucleon mass, $w_t(\z)$ is a noise  with zero mean ($\mathbb E[w_t(\z)]=0$) and correlator:
\bq\label{noisemain}
\mathbb E[w_t(\z)w_s(\x)]=\frac{1}{2\pi}\int_{-\infty}^{+\infty}\D\omega\,\gamma(\omega)e^{-i\omega(t-s)}\delta(\x-\z),\\
\eq
where $\gamma(\omega)=\gamma(-\omega)$ is the frequency-dependent collapse strength. We denoted with $\hat \x_A$ the position operator of the particle, and:
\bq
g(\x)=\frac{e^{-\x^2/2\rC^2}}{(\sqrt{2\pi}\rC)^3}=\frac{1}{(2\pi)^3}\int\D\q\,e^{-\q^2\rC^2/2-i\q\cdot\x}.
\eq
We  assume that the particle is free and confined  in a box of side $L$; the initial and final states read:
\bq
\braket{\x|i}=\frac{e^{i\k_i\cdot\x}}{L^{3/2}},\quad\text{and}\quad
\braket{\x|f}=\frac{e^{i\k_f\cdot\x}}{L^{3/2}}.
\eq
We then have:
\bqali
c_{fi}(t)=&\frac{im_A}{m_0L^3} \int\D\q e^{-\q^2\rC^2/2}\int_0^t\D s\,e^{i\omega_{fi}s}\\
&\times \int\D\z\,w_s(\z)e^{-i\q\cdot\z}\delta(\k_{f}-\k_{i}-\q),
\eqali
where $\omega_{fi}=\omega_f-\omega_i$ and $\k_{i}$, $\k_{f}$ are the initial and final momenta of the particle, respectively. The transition probability, under stochastic average, is then given by
\begin{widetext}
\bq \label{eq:dfgf}
\mathbb E[|c_{fi}|^2]=\frac{m_A^2}{m_0^2L^3}\int\D\q\, e^{-\q^2\rC^2}\int\D\omega\,\gamma(\omega)
\delta(\k_f-\k_i-\q)\,t\,\delta^{(t)}(\omega_{fi}-\omega),
\eq
\end{widetext}
where we used the relations:
\bqali\label{deltat}
\left[\delta(\k_f-\k_i-\q)\right]^2&\sim(L/(2\pi))^3\delta(\k_f-\k_i-\q),\\
\int_0^t\D s\,e^{i(\omega_{fi}-\omega)s}&=2\pi e^{i(\omega_{fi}-\omega)t/2}\delta^{(t)}(\omega_{fi}-\omega),\\
\left[\delta^{(t)}(\omega_{fi}-\omega)\right]^2&\sim (t/(2\pi))\,\delta^{(t)}(\omega_{fi}-\omega).
\eqali

We now apply Eq.~\eqref{eq:dfgf} to the system under study, i.e. a particle in a Fermi gas. The heating power $P_\text{\tiny CSL}(t)=\D E_\text{\tiny TOT}(t)/\D t$  reads:
\bq
P_\text{\tiny CSL}(t)=\frac{\D}{\D t}\sum_i\sum_f\mathcal N(\k_i)\left(1-\mathcal N(\k_f)\right)\hbar\omega_{fi}\mathbb E|c_{fi}(t)|^2,
\eq
where $N(\k_i)$ is the probability of the initial state having momentum $\k_i$, and $\left(1-\mathcal N(\k_f)\right)$ is the probability for the final state with momentum $\k_f$ not to be occupied, otherwise the particle could not jump there because of the Pauli exclusion principle.
Since $\mathcal N(\k_i)\mathcal N(\k_f)$ and $\mathbb E[|c_{fi}|^2]$ are even, whereas $\omega_{fi}$ is odd, under the interchange $i \leftrightarrow f$, the term 
containing $\mathcal N(\k_i)\mathcal N(\k_f)$ makes a vanishing contribution to Eq. (9).  
The above expression then simplifies to
\bq
P_\text{\tiny CSL}(t)=\frac{\D}{\D t}\sum_i\sum_f\mathcal N(\k_i)\hbar\omega_{fi}\mathbb E|c_{fi}(t)|^2.
\eq
Using the standard box-normalization prescription, according to which in the limit $L\to+\infty$:
\bq
\frac{1}{L^3}\sum_\p g(\p)\to\frac{1}{(2\pi)^3}\int\D\p\, g(\p),
\eq
one obtains 
\bq
P_\text{\tiny CSL}(t)=\frac{L^3}{(2\pi)^3}\frac{\D}{\D t}\sum_i\mathcal N(\k_i)\int\D\k_f\,\hbar\omega_{fi}\mathbb E|c_{fi}(t)|^2,
\eq
which in the long time limit reads
\begin{equation}
P_\text{\tiny CSL}(t)=\frac{m_A^2}{m_0^2(2 \pi)^3}\sum_i\,\mathcal N(\k_i)\int\D\q\,\hbar\bar\omega_{i}(\q)e^{-\q^2\rC^2}\gamma(\bar\omega_{i}(\q)),
\end{equation}
where
\begin{equation}
\bar\omega_{i}(\q)=\frac{\hbar}{2m_A}(\q^2+2\k_i\cdot\q).
\end{equation}
In the white noise case, where $\gamma(\omega)=\gamma$, the integration over $\q$  can be easily performed, giving:
\begin{equation}\label{Pcsl}
P_\text{\tiny CSL}(t)=\frac34\frac{\hbar^2\lambda m_A}{m_0^2\rC^2},
\end{equation}
where we used $\gamma=\lambda(\sqrt{4\pi}\rC)^3$ and $\sum_i\,\mathcal N(\k_i)=1$.  
For the $N$ atom case, the calculation of Appendix \ref{app.N} shows that $m_A$ in Eq.~\eqref{Pcsl} is replaced by the total mass $M=N m_A$. This is the same result obtained from the study of phononic modes in matter \cite{Adler:2005ab,Adler:2018aa,Bahrami:2018aa}.

\section{Neutron Stars}

Neutron stars are small (radius $\sim10\,$km) and dense (mass $M\sim1.4-4.2\times10^{30}\,$kg and density $\mu\sim10^{17}\,$kg/m$^3$), resulting from the collapsed cores of stars with mass above the Chandrasekhar limit \cite{Weinberg:1972aa}. After a first stage next to their formation, where they cool through emission of baryonic matter, the main cooling process is dominated by thermal emission of radiation \cite{Lattimer:1994aa,Lattimer:2001aa}, which is described by the Stefan-Boltzmann law:
\bq
P_\text{\tiny rad}=S \sigma T^4,
\eq
where $S$ is the surface of the neutron star, $\sigma=5.6 \times 10^{-8}\,$W m$^{-2}$K$^{-4}$ is the Stefan's constant and $T$ is the effective black-body temperature of the  star. As a reference value for the temperature we can consider $T=0.28^{+0.19}_{-0.12}\times10^6\,$K, which refers to the neutron star \verb|PSR J 1840-1419| \cite{Keane:2013aa}. The radius is  $R=10\,$km and the mass $M=2\times 10^{30}\,$kg, equal to the solar mass, giving a density $\mu=4.8\times 10^{17}\,$kg/m$^3$. Variation of $R$ and $M$, for typical dimensions of a neutron star, do not imply significant changes in the bounds on the CSL parameters.
\begin{figure}[t!]
\centering
\includegraphics[width=\linewidth]{{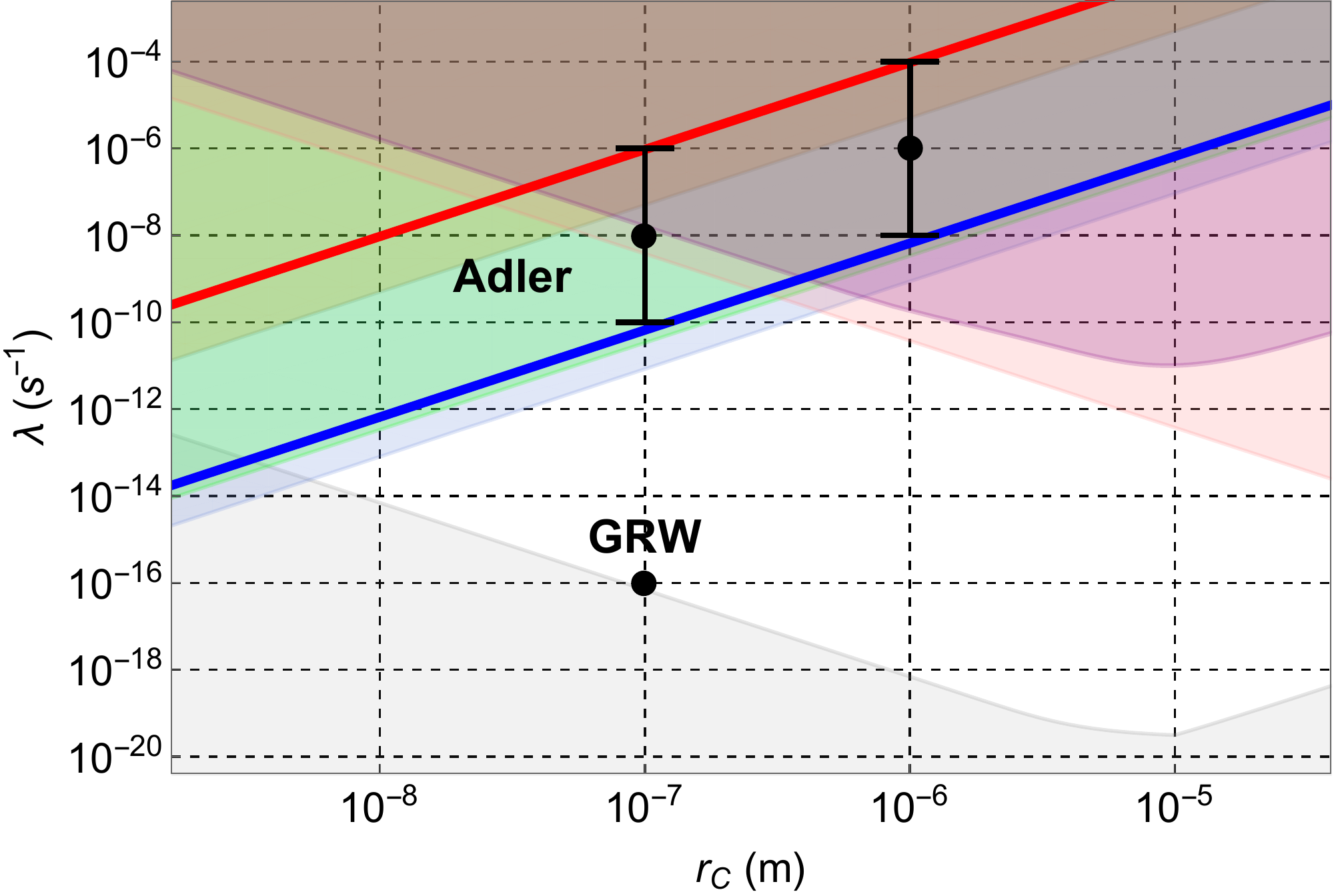}}
\cprotect\caption{\label{fig1}Bounds on the collapse parameters $\lambda$ and $\rC$ for the standard (white noise) mass-proportional CSL model. The red and blue lines denote the upper bounds given by Eq.~\eqref{reslambda} applied to the heat flow from the neutron star \verb|PSR J 1840-1419| and from Neptune. The shaded areas show the already experimentally and theoretically excluded regions: the orange region comes from cold atom experiment \cite{Kovachy:2015ab,Bilardello:2016aa}; the green region from phonon analysis in cryogenic experiments \cite{Pobell:2007aa,Adler:2018aa,Bahrami:2018aa}; the blue region from x-ray emission from germanium \cite{Fu:1997aa,Adler:2013aa,Bassi:2014aa,Donadi:2014aa,Piscicchia:2017aa}; the purple region from mechanical cantilever \cite{Vinante:2016aa,Vinante:2017aa}; the pink region from LISA Pathfinder \cite{Armano:2016aa,Carlesso:2016ac,Armano:2018aa,Carlesso:2018ab}; the grey region from theoretical arguments \cite{Toros:2017aa,Toros:2018aa}. }
\end{figure}

\section{Results and Discussion}

Assuming that the neutron star's thermal radiation emission is balanced by the heating effect due to the CSL noise, we impose
$
P_{\text{\tiny rad}}=P_\text{\tiny CSL}.
$
This gives an estimate of collapse rate:
\bq\label{reslambda}
\lambda = \frac{16 R^2 m_0^2 \pi \rC^2 T^4\sigma}{3M \hbar^2},
\eq
where we assumed that the neutron star can be approximated by a sphere of radius $R$.
The corresponding upper bound is shown in red in Fig.~\ref{fig1}.

It is interesting to apply Eq.~\eqref{reslambda} to other objects in the Universe.
 Table \ref{table1} shows the values of the ratio $P_{\text{\tiny rad}}/M$ and the corresponding value of $\lambda/\rC^2$ for the planets in the Solar system, the Moon, the Sun and, as a comparison, that of the neutron star \verb|PSR J 1840-1419| analized above.
Numbers show that Neptune gives the best ratio $\lambda/\rC^2$, which is more than 4 orders smaller than the neutron star's one. The corresponding upper bound is identified by continuous blue line in Fig.~\ref{fig1}.
These bounds are weaker than the already existing bounds, and are further weakened if one assumes a high-frequency cut off in the noise spectrum following the methods of \cite{Bassi:2009aa,Ferialdi:2012aa,Ferialdi:2012ab,Adler:2018aa,Carlesso:2018aa}, or dissipative modification of the CSL model as shown in \cite{Smirne:2015aa,Bilardello:2016aa,Toros:2017aa,Nobakht:2018aa}.  

\FloatBarrier
\begin{table}[t!]
\cprotect\caption{\label{table1} Numerical values of the ratio $P_{\text{\tiny rad}}/M$ for the planets in the Solar system (Sun and Moon included) \cite{Williams:aa}, and the corresponding value of $\lambda/\rC^2$ according to Eq.~\eqref{reslambda}. For completeness, we report also the values for the neutron star \verb|PSR J 1840-1419| analized above.}
\begin{tabular}{l|ccc}
\hline\hline
&$P_\text{\tiny rad}/M$[W/kg]&$\lambda/\rC^2$[s$^{-1}$m$^{-2}$] \\ 
 \hline
Mercury&$4.74\times10^{-7}$&$1.57\times10^{8}$\\
Venus&$1.40\times10^{-8}$&$4.62\times10^6$\\
Earth&$2.00\times10^{-8}$&$6.60\times10^6$\\
Moon&$1.55\times10^{-7}$&$5.12\times10^7$\\
Mars&$2.45\times10^{-8}$&$8.10\times10^6$\\
Jupiter&$2.76\times10^{-10}$&$9.14\times10^4$\\
Saturn&$1.94\times10^{-10}$&$6.40\times10^4$\\
Uranus&$6.03\times10^{-11}$&$2.00\times10^4$\\
Neptune&$1.99\times10^{-11}$&$6.57\times10^3$\\
Pluto&$1.50\times10^{-10}$&$4.98\times10^4$\\
Sun&$1.90\times10^{-4}$&$6.29\times10^{10}$\\
\hline
Neutron star &$2.85\times10^{-7}$&$9.43\times10^7$\\
\hline\hline
\end{tabular}
\end{table}
\FloatBarrier

\subsection*{Acknowledgments}
SLA acknowledges the hospitality of the Aspen Center for Physics, which is supported 
by the National Science Foundation grant PHY-1607611. AB acknowledges financial support from the COST Action QTSpace (CA15220), INFN and the University of Trieste. AB, MC and AV acknowledge financial support from the H2020 FET Project TEQ (grant n.~766900).

\vfill

   \onecolumngrid
\appendix

\section{Field-theoretical calculation}
\label{app.N}

We perform the same analysis presented in the main text, within the framework of quantum field theory.  Let us consider the CSL Hamiltonian:
\bq
\hat H=\hat H_0+\hat V_\text{\tiny CSL},
\eq
where
\bq
\hat H_0=\sum_i\sum_\tau\sum_\p E_{\p\tau i}\hat b^\dag_{\p\tau i}(t)\hat b_{\p\tau i}(t),
\eq
is the free Hamiltonian; the first sum is over the \textit{i}-type of particle, the second sum over the spin $\tau$ ($i$-th type of particle) and the third over momentum. 
{}{Here $\hat b^\dag_{\p\tau i}$ and $\hat b_{\p\tau i}$ are creation and annihilation operators respectively: since the final result is independent from the particle nature, they can be fermionic or bosonic. In fact, the derivation presented below depends only on the following commutation relations $\com{\hat b^\dag_{\p\tau i}\hat b_{\p'\tau j}}{\hat b^\dag_{\k\tau' l}}=\delta^{(3)}(\p'-\k)\delta_{\tau\tau'}\delta_{jl}\hat b^\dag_{\p\tau i}$ and $\com{\hat b^\dag_{\p\tau i}\hat b_{\p'\tau j}}{\hat b_{\k\tau' l}}=-\delta^{(3)}(\p'-\k)\delta_{\tau\tau'}\delta_{jl}\hat b_{\p\tau i}$, which are identical for fermions and bosons.}
The CSL stochastic potential is \cite{Adler:2013aa}:
\bq
\hat V_\text{\tiny CSL}=-\hbar\sum_j\sum_{\tau'}\frac{m_j}{m_0}\int\D\x\,\hat \Psi_{\tau' j}^\dag(\x,t)\hat \Psi_{\tau' j}(\x,t)\xi(\x,t),
\eq
Here we introduced:
\bq
\xi(\x,t)=\int\D\y\,\frac{e^{-(\x-\y)^2/2\rC^2}}{(\sqrt{2\pi}\rC)^3}w_t(\y),
\eq
whose mean and correlator are:
\bq\label{corrnoise}
\mathbb E[\xi(\x,t)]=0,\quad\text{and}\quad
\mathbb E[\xi(\x,t)\xi(\y,s)]=\tilde\gamma(t-s)F(\x-\y),
\eq
where $\mathbb E$ denotes the stochastic average over the noise, 
\bq\label{noiseapp}
F(\x)=\frac{e^{-\x^2/4\rC^2}}{(\sqrt{4\pi}\rC)^3},\quad\text{and}\quad\tilde\gamma(t)=\frac{1}{2	\pi}\int\D\omega\,\gamma(\omega)e^{-i\omega t}.
\eq
The relation between the operator $\hat \Psi_{\tau j}(\x,t)$ and $\hat b_{\p\tau i}(t)$ is given by
\bqali
&\hat \Psi_{\tau j}(\x,t)=\sum_\p\psi_{\p\tau j}(\x)\hat b_{\p\tau j}(t),\\
&\hat b_{\p\tau j}(t)=\int\D\x\,\psi^*_{\p\tau j}(\x)\hat \Psi_{\tau j}(\x,t),
\eqali
with $\psi_{\p\tau j}(\x)$ denoting the Fourier coefficients of the transformation, spin $\tau$ and of momentum $\p$. Below we specify the exact form of $\psi_{\p\tau j}(\x)$.
The evolution of $\hat b_{\p\tau i}(t)$ is determined by the Heisenberg equation $\frac{\D\hat b_{\p\tau j}(t)}{\D t}=\tfrac i\hbar\com{\hat H}{\hat b_{\p\tau j}(t)}$, which gives
\bqali
\frac{\D\hat b_{\p\tau j}(t)}{\D t}=-\frac i\hbar E_{\p\tau j}\hat b_{\p\tau j}(t)+
i\frac{m_j}{m_0}\sum_\k\int\D\x\,\psi^*_{\p\tau j}(\x)\psi_{\k\tau j}(\x)\xi(\x,t)\hat b_{\k\tau j}(t).
\eqali
The solution is:
\bq\label{solutionB}
\hat b_{\p\tau j}(t)=e^{-\tfrac i\hbar E_{\p \tau j}t}\hat b_{\p\tau j}(0)+i\frac{m_j}{m_0}\sum_\k\int\D\x\,\psi^*_{\p\tau j}(\x)\psi_{\k\tau j}(\x)\int_0^t\D s\,e^{-\tfrac i\hbar E_{\p \tau j}(t-s)}\xi(\x,s)\hat b_{\k\tau j}(s).
\eq
Since $\hat b_{\k\tau j}(s)$ appears also in the last term, we need to solve perturbatively. We replace $\hat b_{\k\tau j}(s)$ with the corresponding form given again by Eq.~\eqref{solutionB}, and truncate the expression to order $\gamma$:
\bq
\hat b_{\p\tau j}(t)=\hat A_{\p\tau j}(t)+\hat B_{\p\tau j}(t)+\hat C_{\p\tau j}(t)+\mathcal O(\gamma^{3/2}),
\eq
where
\bqali\label{defABC}
\hat A_{\p\tau j}(t)&=e^{-\tfrac i\hbar E_{\p \tau j}t}\hat b_{\p\tau j}(0),\\
\hat B_{\p\tau j}(t)&=i\frac{m_j}{m_0}\sum_\k\int\D\x\,\psi^*_{\p\tau j}(\x)\psi_{\k\tau j}(\x)\int_0^t\D s\,e^{-\tfrac i\hbar E_{\p \tau j}(t-s)}\xi(\x,s)e^{-\tfrac i\hbar E_{\k \tau j}s}\hat b_{\k\tau j}(0),\\
\hat C_{\p\tau j}(t)&=-\left(\frac{m_j}{m_0}\right)^2\sum_{\k\k'}\int\D\x\,\psi^*_{\p\tau j}(\x)\psi_{\k\tau j}(\x)\int_0^t\D s\,e^{-\tfrac i\hbar E_{\p \tau j}(t-s)}\xi(\x,s)\int\D\y\,\psi^*_{\k\tau j}(\y)\psi_{\k'\tau j}(\y)\times\\
&\int_0^s\D s'\,e^{-\tfrac i\hbar E_{\k \tau j}(s-s')}\xi(\y,s')
e^{-\tfrac i\hbar E_{\k' \tau j}s'}\hat b_{\k'\tau j}(0).
\eqali
Given these expressions, we can compute the evolution of the energy expectation value, which is given by
\bq
E_\text{\tiny TOT}(t)=\mathbb E[\braket{\hat H}].
\eq
Due to the stochastic properties in Eq.~\eqref{corrnoise}, we have $\mathbb E[\hat V_\text{\tiny CSL}]=0$, therefore only $\hat H_0$ contributes to $E_\text{\tiny TOT}(t)$. In particular
\bq
E_\text{\tiny TOT}(t)=E_\text{\tiny TOT}(0)+E_\text{\tiny TOT}^\text{\tiny CSL,1}(t)+E_\text{\tiny TOT}^\text{\tiny CSL,2}(t)+\mathcal O(\gamma^{3/2}),
\eq
where
\bqali
E_\text{\tiny TOT}(0)&=\sum_i\sum_\tau\sum_\p E_{\p\tau i}\braket{\hat A^\dag_{\p\tau i}(t)\hat A_{\p\tau i}(t)},\\
E_\text{\tiny TOT}^\text{\tiny CSL,1}(t)&=\sum_i\sum_\tau\sum_\p E_{\p\tau i}\braket{\hat B^\dag_{\p\tau i}(t)\hat B_{\p\tau i}(t)},\\
E_\text{\tiny TOT}^\text{\tiny CSL,2}(t)&=\sum_i\sum_\tau\sum_\p E_{\p\tau i}\braket{\hat A^\dag_{\p\tau i}(t)\hat C_{\p\tau i}(t)+\text{H.C.}},
\eqali
where H.C. stands for hermitian conjugate. We notice that there is no contribution from terms like $\hat A^\dag_{\p\tau i}(t)\hat B_{\p\tau i}(t)$ or $\hat B^\dag_{\p\tau i}(t)\hat C_{\p\tau i}(t)$: the first is zero under stochastic average and the second  scales with $\gamma^{3/2}$ and can be then neglected. The above expressions, together with Eq.~\eqref{defABC}, give:
\bqali
E_\text{\tiny TOT}(0)&=\sum_i\sum_\tau\sum_\p E_{\p\tau i}\braket{\hat b^\dag_{\p\tau i}(0)\hat b_{\p\tau i}(0)},\\
E_\text{\tiny TOT}^\text{\tiny CSL,1}(t)&=\sum_i\sum_\tau\sum_\p E_{\p\tau i}\left(\frac{m_i}{m_0}\right)^2\sum_{\k\k'}\int\D\x\int\D\y\,\psi_{\p\tau i}(\x)\psi^*_{\k\tau i}(\x)\psi^*_{\p\tau i}(\y)\psi_{\k'\tau i}(\y)F(\x-\y)\times\\
&\quad\int_0^t\D s\int_0^t\D s'\, \tilde\gamma(s-s')e^{-\tfrac i\hbar E_{\p \tau i}(s-s')}e^{\tfrac i\hbar E_{\k \tau i}s}e^{-\tfrac i\hbar E_{\k' \tau i}s'}
\braket{\hat b^\dag_{\k\tau i}(0)b_{\k'\tau i}(0)},\\
E_\text{\tiny TOT}^\text{\tiny CSL,2}(t)&=-\sum_i\sum_\tau\sum_{\p} E_{\p\tau i}\left(\frac{m_i}{m_0}\right)^2\sum_{\k\k'}
\int\D\x\int\D\y\,F(\x-\y)\psi_{\k \tau i}(\x)\psi^*_{\k \tau i}(\y)\int_0^t\D s\int_0^s\D s'\,\tilde\gamma(s-s')\times\\
&\left[\psi^*_{\p\tau i}(\x)\psi_{\k' \tau i}(\y)e^{\tfrac i\hbar E_{\p \tau i}s}e^{-\tfrac i\hbar E_{\k \tau i}(s-s')}e^{-\tfrac i\hbar E_{\k' \tau i}s'}\braket{\hat b^\dag_{\p \tau i}(0)\hat b_{\k'\tau i}(0)}+\right.\\
&\left.\quad\psi_{\p\tau i}(\y)\psi^*_{\k' \tau i}(\x)e^{-\tfrac i\hbar E_{\p \tau i}s}e^{\tfrac i\hbar E_{\k \tau i}(s-s')}e^{\tfrac i\hbar E_{\k' \tau i}s'}\braket{\hat b^\dag_{\k' \tau i}(0)\hat b_{\p\tau i}(0)}\right].
\eqali
The above terms contain $\braket{\hat b^\dag_{\p \tau i}(0)\hat b_{\k\tau i}(0)}$. To compute it we consider a state of $N$ particles with density matrix diagonal in momentum and weight given by the occupation number $\mathcal N(\p)$. Then we have
\bq
\braket{\hat b^\dag_{\p \tau i}(0)\hat b_{\k\tau i}(0)}=\delta_{\p\k}\mathcal N(\p).
\eq
 {Although $\mathcal N(\p)$ is different in the fermionic and the bosonic case, as it should be clear from the calculations, the final result is independent from the type of statistics.} 
Applying this result we obtain
\bqali
E_\text{\tiny TOT}(0)&=\sum_i\sum_\tau\sum_\p E_{\p\tau i}\mathcal N(\p),\\
E_\text{\tiny TOT}^\text{\tiny CSL,1}(t)&=t\sum_i\sum_\tau\sum_{\p\k} E_{\p\tau i}\left(\frac{m_i}{m_0}\right)^2\mathcal N(\k)
\int\D\x\int\D\y\,\psi_{\p\tau i}(\x)\psi^*_{\k\tau i}(\x)\psi^*_{\p\tau i}(\y)\psi_{\k\tau i}(\y)F(\x-\y)\times\\
& \int\D\omega\,\gamma(\omega)\,\delta^{(t)}\left(\tfrac{E_{\p\tau i}-E_{\k\tau i}}{\hbar}-\omega\right),\\
E_\text{\tiny TOT}^\text{\tiny CSL,2}(t)&=-t\sum_i\sum_\tau\sum_{\p,\k} E_{\p\tau i}\mathcal N(\p)\left(\frac{m_i}{m_0}\right)^2
\int\D\x\int\D\y\,\psi_{\k \tau i}(\x)\psi^*_{\p\tau i}(\x)\psi_{\p \tau i}(\y)\psi^*_{\k \tau i}(\y)F(\x-\y)\times\\
& \int\D\omega\,\gamma(\omega)\,\delta^{(t)}\left(\tfrac{E_{\p\tau i}-E_{\k\tau i}}{\hbar}-\omega\right)
\eqali
where we exploited the relations in Eq.~\eqref{deltat} and Eq.~\eqref{noiseapp}.

So far the result is general. We now apply it to the case of interest, i.e. $N$ particles in a cube box of length $L$. We apply the periodic boundary conditions and the box-normalization prescription
\bq
\psi_{\p\tau i}(\x)\to\phi_{\q \tau i}(\x)=\frac{e^{i\q_{\tau i}\cdot\x}}{L^{3/2}},\quad\text{with}\quad\q_{\tau i}=\frac{2\pi}{L}\n_{\tau i},
\eq
where $\n_{\tau i}\in \mathbb Z^3$. The wavefunctions $\phi_{\q \tau i}(\x)$ are orthonormal
\bq
\int_{-\tfrac L2}^{\tfrac L2}\int_{-\tfrac L2}^{\tfrac L2}\int_{-\tfrac L2}^{\tfrac L2}\D\x\,\phi_\q(\x)\phi^*_{\q'}(\x)=\delta_{\n\n'}.
\eq
In the $L \rightarrow + \infty$  limit (so that space-integrals extend over the whole space and can be performed exactly) we have
\bqali
E_\text{\tiny TOT}^\text{\tiny CSL,1}(t)&=t\sum_i\sum_\tau\sum_{\p\k} E_{\p\tau i}\left(\frac{m_i}{m_0}\right)^2\mathcal N(\k)\frac{e^{-(\p-\k)^2\rC^2}}{L^3}
 \int\D\omega\,\gamma(\omega)\,\delta^{(t)}\left(\tfrac{E_{\p\tau i}-E_{\k\tau i}}{\hbar}-\omega\right),\\
E_\text{\tiny TOT}^\text{\tiny CSL,2}(t)&=-t\sum_i\sum_\tau\sum_{\p,\k} E_{\p\tau i}\mathcal N(\p)\left(\frac{m_i}{m_0}\right)^2\frac{e^{-(\p-\k)^2\rC^2}}{L^3}
 \int\D\omega\,\gamma(\omega)\,\delta^{(t)}\left(\tfrac{E_{\p\tau i}-E_{\k\tau i}}{\hbar}-\omega\right),
\eqali
The CSL heating power $P_\text{\tiny CSL}=\tfrac{\D}{\D t}E_\text{\tiny TOT}(t)$ in the long time limit is then given by
\bqali\label{eqP1}
P_\text{\tiny CSL}=\sum_i\sum_\tau\sum_\p\left(\frac{m_i}{m_0}\right)^2\mathcal N(\p)
\frac{1}{L^3}\sum_\k e^{-(\p-\k)^2\rC^2}\left(E_{\k\tau i} -E_{\p\tau i}\right)\gamma(\tfrac{E_{\p\tau i}-E_{\k\tau i}}{\hbar}).
\eqali
In the white noise case, where $\gamma(\omega)=\gamma=\lambda(2\sqrt{\pi}\rC)^3$, by taking $E_{\k\tau i}=\hbar^2\k^2/(2m_i)$ we find
\bq
\frac{\gamma}{L^3}\sum_\k  e^{-(\p-\k)^2\rC^2}(E_{\k\tau i}-E_{\p\tau i}) \quad \xrightarrow[L \rightarrow + \infty]{} \quad  \frac{3\hbar^2\lambda}{4m_i\rC^2}.
\eq
By merging with Eq.~\eqref{eqP1} we have:
\bq
P_\text{\tiny CSL}=\frac{3\hbar^2\lambda }{4m_0^2\rC^2}\sum_im_i\sum_\tau\sum_\p\mathcal N(\p)=\frac{3\hbar^2\lambda M}{4m_0^2\rC^2},
\eq
since that $\sum_\tau\sum_\p\mathcal N(\p)$ gives the number of particle of type $i$.

\end{document}